\documentclass[11pt]{article}
\usepackage{amsmath,amssymb,color,graphics,epsfig,cite}

\usepackage{amsfonts}

\newcommand{\be}{\begin{equation}}
\newcommand{\ee}{\end{equation}}
\newcommand{\bea}{\setlength\arraycolsep{2pt} \begin{eqnarray}}
\newcommand{\eea}{\end{eqnarray}}
\newcommand{\nn}{\nonumber}

\def\ft#1#2{{\textstyle{\frac{\scriptstyle #1}{\scriptstyle #2} } }}
\def\fft#1#2{{\frac{#1}{#2}}}

\def\0{{\sst{(0)}}}
\def\1{{\sst{(1)}}}
\def\2{{\sst{(2)}}}
\def\3{{\sst{(3)}}}
\def\4{{\sst{(4)}}}
\def\5{{\sst{(5)}}}
\def\6{{\sst{(6)}}}
\def\7{{\sst{(7)}}}
\def\8{{\sst{(8)}}}
\def\sst#1{{\scriptscriptstyle #1}}

\begin{document}

\begin{center}
{\Large {\bf General Solutions of Einstein Gravity at $D\rightarrow 2$}}

\vspace{20pt}

Qi-Yuan Mao and H. L\"{u}

\vspace{10pt}

{\it Center for Joint Quantum Studies and Department of Physics,\\
School of Science, Tianjin University, Tianjin 300350, China }

\vspace{40pt}

\underline{ABSTRACT}
\end{center}

Einstein gravity at $D\rightarrow 2$ limit can be obtained from the Kaluza-Klein procedure by taking the dimensions of the internal space to zero while keeping only the breathing mode. The resulting scalar-tensor theory can be further reduced to JT gravity or the Liouville CFT at the large central charge limit, bridging the two important 2d models. We study the general solutions of the theory, including black holes and wormholes for both positive and negative cosmological constants. We obtain the on-shell action of the nearly AdS$_2$ and show that for suitable boundary slices, the Schwarzian action governs the leading-order dynamics at the finite boundary cutoff in later time. For positive cosmological constant, we find that the scalar is well defined on the 2-sphere.



\thispagestyle{empty}
\pagebreak

\tableofcontents
\addtocontents{toc}{\protect\setcounter{tocdepth}{2}}


\section{Introduction}
\label{sec:intr}

Two-dimensional (2d) gravity was well studied in the 80s owing to its close connection to the perturbative string theory \cite{Polyakov:1981rd,Ginsparg:1993is}. Notable examples include Liouville conformal field theory (CFT) \cite{Polyakov:1981rd,DHoker:1982wmk}. There has been a resurgence in the topic in the context of the anti-de Sitter(AdS)/CFT correspondence. It was proposed \cite{Maldacena:2016upp} that JT gravity \cite{Teitelboim:1983ux,Jackiw:1984je} may provide a gravity dual to the SKY model \cite{Sachdev:1992fk,kitaevsyk}. Recently, the work \cite{Almheiri:2019psf, Almheiri:2019qdq} on Page curve \cite{Page:1993wv} of evaporating black hole of JT gravity has brought much attention to 2d gravitational models. Owing to the fact that the Einstein-Hilbert action in two dimensions has no dynamics but describes the topology of 2d manifolds, 2d models are either beyond Einstein gravity, such as the Liouville CFT, or simply the Kaluza-Klein reduction of Einstein gravity in higher dimensions, such as JT gravity.

If one would like to define a theory based on its characteristic solutions, Einstein gravity in two dimensions ought to exist, since the Schwarzschild-AdS black hole has a straightforward and smooth $D\rightarrow 2$ limit. (Its $D\rightarrow 3$ limit is in fact much more subtle in that the naively expected logarithmic dependence is actually absent, as we can see in the well-known BTZ black hole \cite{Banados:1992wn}.) We can refer to such a 2d model as Einstein gravity at $D\rightarrow 2$ limit.  The theory was first constructed in \cite{Mann:1992ar}, by taking the conformal transformation of Einstein gravity in general $D$ dimensions that introduces a scalar field. A nontrivial scalar-tensor theory emerges after taking an appropriate $D\rightarrow 2$ limit. It was later observed \cite{Grumiller:2007wb} that the theory can also be obtained from the Kaluza-Klein approach by taking the zero limit of the dimensions of the internal space. Specifically, one may consider Einstein gravity in general $D=n+2$ dimensions, reduced on some $n$ dimensional internal space, keeping only the breathing zero mode \cite{Grumiller:2007wb}:
\be
d\hat s_{n+2}^2 = ds_2^2 + e^{2\phi} d\Sigma_n^2\,.\label{reduction}
\ee
Here $d\Sigma_n^2$ is a compact Einstein metric with constant curvature $\lambda$: $\tilde R_{ij}=(n-1) \lambda \tilde g_{ij}$ and $\tilde R=n(n-1)\lambda $. After taking an appropriate $n\rightarrow 0$ limit, one not only reproduces the theory of \cite{Mann:1992ar}, but also introduces a Liouville-like exponential potential $\lambda e^{-2\phi}$. Indeed this limit of Einstein gravity at $D\rightarrow 2$ gives rise to the Liouville CFT at $c\rightarrow \infty$ limit, where $c$ is the central charge \cite{Grumiller:2007wb}. This procedure was later independently developed \cite{Lu:2020iav,Kobayashi:2020wqy} to obtain a Horndeski-like scalar-tensor theory in four dimensions as a candidate for the Einstein-Gauss-Bonnet theory at $D\rightarrow 4$ limit, proposed in \cite{Glavan:2019inb}. The generalization to general even dimensions were given in \cite{Matsumoto:2022fln}. As in the case of pure gravities in general dimensions \cite{Mukhopadhyay:2006vu,Khodabakhshi:2022zye}, these special scalar-tensor theories also have Lagrangian holographic relations \cite{Khodabakhshi:2022knu}.

We shall review Einstein gravity at $D\rightarrow 2$ limit in section 2. Depending whether we include a minimally-coupled cosmological constant, the theory will not only lead to the Liouville CFT at large $c$, but also to JT gravity, making it a particularly interesting 2d model. In section 3, we construct both the most general static and time-dependent AdS$_2$ black holes and study their global structures. As in the case of JT gravity, the curvature tensor is constant and we obtain the most general scalar profile and study how the scalar hair affects the black hole thermodynamics and stability. Unlike JT gravity, Einstein gravity at $D\rightarrow 2$ can also have a scalar fixed point such that the vacuum preserves the full AdS$_2$ isometry. In section 4, we consider solutions with positive cosmological constant. In this case, the solutions naturally describe traversable wormholes connecting two cosmic horizons.  In sections 5 and 6, we study the Euclidean-signatured nearly AdS$_2$ space, which can be either hyperbolic 2-space if the cosmological constant is negative, or 2-sphere if the cosmological constant is positive. In the latter case, the scalar is well defined on the 2-sphere. We obtain the on-shell actions and study whether the Schwarzian action can emerge in the former case. We conclude the paper in section 7.

\section{The theory}
\label{sec:theory}

\subsection{A $D\rightarrow 2$ limit of Einstein gravity}

The action of Einstein gravity coupled to a cosmological constant $\Lambda=-\fft12 n(n+1)/\ell^2$, together with some minimally coupled matter in general $D=n+2$ dimensions takes the form
\bea
S&=& S_g + S_m\,,\qquad S_m = \int d^{n+2} x\,{\cal L}_m\,,\nn\\
S_g &=& \fft{1}{16\pi G} \int d^{n+2} x\,\sqrt{-g}\Big( R + \fft{n(n+1)}{\ell^2}\Big).
\eea
It is consistent to perform Kaluza-Klein reduction on some Einstein spaces, keeping only the breathing scalar mode $\phi$. The reduction ansatz is given by \eqref{reduction}. For the gravity sector $S_g$, the resulting 2d Lagrangian is
\be
{\cal L}_2^{(n)}=\sqrt{-g} e^{n\phi} \Big(R+ n(n+1)\ell^{-2}+ n(n-1)\lambda e^{-2\phi} + n(n-1) (\nabla\phi)^2\Big).\label{lagn}
\ee
There are two special cases of particular interest, namely $n=1$ and $n=0$. The former leads to JT gravity \cite{Teitelboim:1983ux,Jackiw:1984je}:
\be
{\cal L}_2^{(1)} = \sqrt{-g} \varphi \big (R + 2 \ell^{-2}\big)\,,
\ee
where we use $\varphi$ to rename $e^{\phi}$ that describes the radius of the $U(1)$ cycle in $D=3$. If we take $n=0$, we arrive at a purely topological term ${\cal L}^{(0)} = \sqrt{-g} R$. However a nontrivial result can nevertheless be obtained in the $n\rightarrow 0$ limit
when the Newton's constant simultaneously goes zero. The resulting scalar-tensor theory can be viewed as Einstein gravity at the $D\rightarrow 2$ limit, and it is this theory we would like to study. The action in the $n\rightarrow 0$ gives
\be
S=\fft{1}{16\pi G} \int d^2x\, \sqrt{-g} R + \fft{n}{16\pi G} \int d^2x\, \sqrt{-g}\Big(\phi R +\fft{1}{\ell^2} -\lambda\, e^{-2\phi} - (\nabla\phi)^2\Big) + \fft{{\cal O}\big(n^2\big)}{16\pi G} \,.\label{2daction}
\ee
The first term is purely topological, giving rise to the Euler number of the 2d manifold; it diverges in the small $G$ limit. The leading-order dynamics is governed by the second term. From the action point of view, the physical degrees of freedom of the 2d gravity describes the small perturbation to the topological Euler number. Setting $\lambda=0$ leads to the theory of \cite{Mann:1992ar}. We shall call the minimally-coupled $\Lambda_0=-1/\ell^2$ as the bare cosmological constant, which can be positive, negative or zero.

\subsection{Limiting to JT gravity}

As we shall see presently, the equations of motion indicate that the curvature of the 2d spacetime is constant, which is a characteristics of JT gravity. However, unlike JT gravity, the theory we obtained has a fixed point for the scalar field, namely
\be
\phi=\bar\phi =\ft12 \log(\lambda \ell^2)\,.\label{vacphi}
\ee
This implies that Einstein gravity at $D\rightarrow 2$ admits a true AdS$_2$ vacuum, and the corresponding dual field theory is conformal with the full Virasoro algebra, rather than some nearly conformal field theory (NCFT) such as the SYK model, where the symmetry is broken into its $SL(2,\mathbb R)$ subgroup \cite{Maldacena:2016hyu}.

We can perform perturbation of this AdS$_2$ vacuum, with the scalar field $\phi=\bar\phi +\epsilon \varphi$, $\epsilon\rightarrow 0$. It is straightforward to see that the leading-order perturbation of \eqref{2daction} is JT gravity, namely
\be
S=\fft{1 + n\bar\phi}{16\pi G} \int d^2x\, \sqrt{-g} R + \fft{n\epsilon}{16\pi G} \int d^2x\, \sqrt{-g}\varphi \Big(R +\fft{2}{\ell^2}\Big) + \cdots.\label{2daction1}
\ee
where the ellipsis denotes the higher-order terms in $n$ and $\epsilon$. We may choose  $\epsilon\sim n$, in which case, we have $G\sim n^2$.  It is rather intriguing that JT gravity from the $n=1$ reduction can also be obtained from some appropriate $n\rightarrow 0$ limit.
It should be pointed out that the CFT$_1$ (in one dimension) is vacuous and trivial, and nontrivial dynamics exists in NCFT$_1$ that is dual to the nearly AdS$_2$ where the scalar field is turned on. The fact that JT gravity can emerge indicates that the theory is nontrivial and it provides a connection between CFT$_1$ and NCFT$_1$.

The theory (\ref{2daction}) has a symmetry of constant shift of the scalar field, together with the redefinition of the coupling constant $\lambda$. The effect of the transformation is to introduce a non-dynamic Einstein-Hilbert term. Thus a local $\phi=0$ at certain point in spacetime is not a singularity. In fact in the AdS$_2$ vacuum, the constant scalar $\phi$ can be set zero globally by the shifting symmetry. In the case of JT gravity derived above, on the other hand, this symmetry is fixed at $\phi=\bar \phi$ and hence the local $\varphi=0$ is singular.

\subsection{Limiting to Liouville CFT}

The relation to Liouville theory was observed in \cite{Grumiller:2007wb}. The semi-classical limit of Liouville CFT in a 2d {\it space} is defined by the local action
\be
S=\fft{1}{4\pi} \int d^2 x\, \sqrt{g}\Big(Q\varphi R +(\partial\varphi)^2 + \tilde\lambda\, e^{2b\varphi}\Big)\,,
\ee
where the parameter $Q$ is called the background charge. The conformal invariance of the action requires what
\be
Q=b + \fft{1}{b}\,.
\ee
The corresponding central charge of the Virasoro algebra is $c=1 + 6Q^2$. At the first sight, the theory \eqref{2daction} does not fit the pattern. We can redefine the scalar by
\be
\phi = -b \varphi\,,\qquad b=\sqrt{\fft{4G}{n}}\,,
\ee
The action \eqref{2daction} becomes
\be
S=\fft1{16\pi G} \int d^2 x \sqrt{|g|} R - \fft{1}{4\pi} \int d^2 x \sqrt{|g|}\Big(
\fft{1}{b} \varphi R - \fft{1}{\tilde\ell^2}+ (\partial\varphi)^2 + \tilde\lambda e^{2b\varphi}\Big) + \cdots\,.\label{2daction2}
\ee
where $\tilde\lambda=\lambda/b^2$ and $\tilde \ell=b\ell$.
After setting $1/\tilde\ell=0$, this theory can match the semiclassical limit of Liouville CFT, provided that $b\ll 1$, in which case, we have
\be
Q=b+\fft{1}{b}\sim \fft{1}{b} \rightarrow \infty\,,\qquad\longrightarrow\qquad c\sim \fft{6}{b^2}\rightarrow \infty\,.
\ee
The overall minus sign in the second term of \eqref{2daction2} is due to the fact that the theory is in the Lorentzian signature. Thus the Liouville theory with large central charge arises from Einstein gravity with zero bare cosmological constant in two dimensions under the limit
\be
G\ll n \ll 1\,.
\ee
One way to impose the above inequality constraint is to let $G\propto n^2$. We then have $b\propto \sqrt{n}$, $Q\propto 1/\sqrt{n}$ and $c\propto 1/n$, in the limit of $n\rightarrow 0$.

Thus we see that in the same $G\sim n^2\rightarrow 0$ limit, both JT gravity and the Liouville theory can arise; the former requires a nonvanishing bare cosmological constant $\Lambda_0=-1/\ell^2$ such that the scalar has a fixed point, and JT gravity describes the perturbative dynamics of this fixed point. The latter requires vanishing $\Lambda_0$ and hence the 2d spacetime is flat. Einstein gravity at $D\rightarrow 2$ limit thus bridges JT gravity and Liouville CFT at large $c$.

\subsection{Equations of motion}

We have so far considered the dimensional reduction only of the gravity sector. If the minimally-coupled matter sector involves only the scalar and vector fields, its Lagrangian ${\cal L}_m$ will simply descend down to $D=2$ in the same form at the $n\rightarrow 0$ limit. We thus have an effect Lagrangian
\be
{\cal L}_{\rm tot}=\sqrt{-g}\Big(\phi R +\fft{1}{\ell^2} -\lambda\, e^{-2\phi} - (\nabla\phi)^2\Big) + {\cal L}_m\,.\label{2dlag}
\ee
The scalar and Einstein equations of motion are given by
\bea
\delta \phi:&& R + 2\lambda {e^{ - 2\phi }}+ 2 \Box \phi  = 0\,,\nn\\
\delta g^{\mu\nu}:&& {g_{\mu \nu }}\square \phi  - {\nabla _\mu }{\nabla _\nu }\phi  - \frac{1}{2}{g_{\mu \nu }}\left( {{\ell^{-2}} - \lambda {e^{ - 2\phi }} - {{\left( {\nabla \phi } \right)}^2}} \right) - {\nabla _\mu }\phi {\nabla _\nu }\phi =  T_{\mu\nu}^{\rm m}\,.\label{geneom0}
\eea
The second equation above can be split into the trace and traceless parts:
\be
\Box \phi={\ell^{-2}} - \lambda{e^{ - 2\phi }} + T^{\rm m}\,,\qquad
\nabla_\mu \phi \nabla_\nu \phi - \ft12 g_{\mu\nu} (\nabla\phi)^2 +
\nabla_\mu\nabla_\nu \phi - \ft12 g_{\mu\nu} \Box\phi=0\,.\label{geneom}
\ee
Together with the scalar equation, we arrive at
\be
R + \fft{2}{\ell^2} = -2T^{\rm m}\,.\label{Reom}
\ee
It is now worth examining how the Schwarzschild-AdS solution, with $T^{\rm m}=0$, reduces to two dimensions. The spherically-symmetric and static solution in general $(n+2)$ dimensions takes the form
\be
ds_{n+2}^2 = -f dt^2 + \fft{dr^2}{f} + r^2 d\Omega_n^2\,,\qquad f=\fft{r^2}{\ell^2} + 1 - \fft{2m}{r^{n-1}}\,.
\ee
It follows from the reduction-based construction, in the limit of $n\rightarrow 0$, we have
\be
ds_2^2 = -f dt^2 + \fft{dr^2}{f}\,,\qquad f = \fft{r^2}{\ell^2} + 1 -2m r\,,\qquad
\phi=\ft12 \log(\lambda r^2)\,,\label{sol0}
\ee
It is easy to verify that the 2d black hole satisfies all the equations in \eqref{geneom0}.

\section{Black holes}
\label{sec:bh}

In this paper, we shall focus on the study of the Lagrangian \eqref{2dlag}, but with no matter fields. We have already obtained a black hole solution \eqref{sol0} that arises from the limiting case of Schwarzschild-AdS black holes in general dimensions. We expect that the theory admits a more general class of black holes involving an independent scalar hair parameter. The 2d theory is relatively simple and we are to construct the most general static and time-dependent black holes. Discussions of black holes in a more general class of theories can be found in \cite{Grumiller:2007ju}.

\subsection{General static black holes}

The equations of motion indicates that the metric must have constant curvature $R=-2/\ell^2$. In two dimensions, this means the metric must be locally AdS$_2$.  We shall take the metric to be the Schwarzschild-AdS, given by
\be
ds_2^2 = - f dt^2 + \fft{dr^2}{f}\,,\qquad f=\fft{r^2}{\ell^2} + k - 2m r\,.
\ee
The $m=0$ and $k=1$ metric is the AdS$_2$ in global coordinates. Note that the linear $r$-dependence in $f$ can be removed by a constant shift of the $r$ coordinate $r\rightarrow r+ m\ell^2$ such that $f\rightarrow f=r^2/\ell^2 + k - m^2\ell^2$, while the metric ansatz remains in the same form. For the solution to be static, we require that the scalar field depends only on the $r$ coordinate. The scalar equation can be solved straightforwardly, giving
\be
\phi = \ft12 \log\big(\lambda \fft{(r-a)^2}{ f(a)}\big)\,,\label{staticphi}
\ee
where $a$ is an independent scalar hair parameter. The black hole \eqref{sol0} corresponds to taking $a=0$ and $k=1$. The $r$-dependence of the scalar field implies that the AdS$_2$ isometry of the metric is broken by the scalar field so that the solution is nearly AdS$_2$. This is analogous to JT gravity; however, the full AdS$_2$ vacuum with constant scalar \eqref{vacphi} exists in our theory. This vacuum solution does not arise from some asymptotic regions, but only emerges if we take parameter $a\rightarrow \pm \infty$ in \eqref{staticphi}.
This should be contrasted with the usual AdS/CFT picture in higher dimensions, where the AdS geometry is typically restored as the radial coordinate $r$ runs from the bulk to asymptotic infinity which is the AdS boundary. This process is the dual version of renormalization group flow to the ultraviolet conformal fixed point. However, in the 2d theory the flow is not described by the radial variable $r$, but by the parameter $a$ associated with the scalar charge that we shall discuss momentarily.

We now analyse the global properties of the local solution. The static solution is invariant under the constant coordinate scaling $t\rightarrow c  t$ and $r\rightarrow c r$, together with the appropriate scaling of the integration constants, namely $(k,m,a)\rightarrow (c^2k,cm,ca)$. This implies that we can set $k=0, \pm 1$, without loss of generality. The solution thus contains two continuous integration constants $m$ and $a$. The metric has in general two horizons $(r_-\le r_+)$:
\be
r_\pm = m\ell^2 \pm \ell \sqrt{m^2\ell^2-k}\,.
\ee
The outer horizon $r_+$ is the event horizon, and corresponding temperature and entropy are
\be
T=\fft{\sqrt{\ell^2 m^2 - k}}{2\pi \ell}\,,\qquad S=\ft14 \phi(r_+)\,.
\ee
Although the metric has no curvature singularity, but the scalar can have. As discussed in section 2, the scalar in our theory is dilaton-like and the dynamics is invariant under the constant shift of the scalar. Therefore $\phi=0$ is not a singularity. The $r\rightarrow \infty$ boundary is not singular either since both $e^{-2\phi}$ and $(\partial\phi)^2$ converge.  The scalar singularity is at $r=a$, where $e^\phi$ vanishes.

For the black hole to be well defined on and outside of the horizon, restriction on the integration constant $a$ must be imposed. Note that $f(a)=(a-r_+)(a-r_-)/\ell^2$.
We therefore must impose $a<r_-$ such that $f(a)>0$. In other words, the parameter $a$ lies in the region
\be
a\in (-\infty, r_-)\,.
\ee
In order to determine the black hole mass, we appeal to the free energy from the Euclidean action, which, including the boundary terms, is given by
\be
16\pi S_{\rm E}=\int_{\cal M} d^2x\,\sqrt{g} \Big(\phi R +\fft{1}{\ell^2} -\lambda\, e^{-2\phi} - (\nabla\phi)^2\Big) + 2\int_{\partial {\cal M}} dy \sqrt{h}(\phi K - \fft{1}{\ell})\,.\label{totalaction0}
\ee
The resulting free energy is
\be
F=\fft{\ell^2 m - a}{8\pi \ell^2} - \fft{\sqrt{\ell^2 m^2 - k}}{8\pi \ell}\phi(r_+)\,.
\ee
This suggests that the mass is not only determined by the metric, but also modified by the parameter $a$ of the scalar hair, namely
\be
M=\fft{\ell^2 m - a}{8\pi \ell^2}\,,
\ee
such that the free energy can be expressed as $F=M - T S$, {\it i.e.}~it is the Helmholtz free energy. That the mass depends on the scalar hair is not surprising for a non-minimally coupled theory of gravity and matter. Since the constant $\phi$ solution arises from the $a\rightarrow -\infty$ limit, it follows that even the AdS$_2$ vacuum with $m=0$, has infinite mass or energy, indicating that it is not stable.

We find that the first law can then be given by
\be
dM= T dS + \Phi dQ\,,\qquad \hbox{with}\qquad
Q =f(a)\,,\qquad \Phi=\fft{M}{2 f(a)}\,,\label{firstlaw}
\ee
The ``charge'' $Q$ associated with the scalar hair can be recognized as the leading coefficient of $e^{-2\phi}$ at the boundary $r\rightarrow \infty$. However, we do not have an independent way of determining its thermodynamic potential $\Phi$, other than by completing the first law, which is in itself quite nontrivial. In this picture, the $a=0$ solution \eqref{sol0} is not hairless, since $f(0)$ does not vanish in general. In fact all black holes necessarily contain scalar hair since $f(a)$ should not vanish.

Since the solution contains additional dimensionful parameters such as $(\lambda, \ell)$, we cannot read off the Smarr relation from the first law. Instead it is simply given by
\be
M=2\Phi Q\,.\label{smarr}
\ee
We may also treat the bare cosmological constant $\Lambda_0=-1/\ell^2$ as a thermodynamic pressure or the force $P=1/(16\pi \ell^2)$, the first law can then be augmented further by $L dP$ where $L=(r_+-a)$ is the black hole ``thermodynamic length'' conjugate to the force. Note that within the allowed parameter region, the length $L$ is positive definite.

Having obtained the black hole thermodynamic variables that satisfy the first law, we can now discuss some of the thermodynamic properties. It is instructive to introduce a dimensionless parameter $b$, defined by
\be
a=r_- - 8 \pi \ell\,b\,.\label{ab}
\ee
The condition $a<r_-$ becomes $b>0$. We shall focus on $k=1$, in which case, the asymptotic AdS$_2$ is in global coordinates. The black hole becomes extremal at $m_{\rm ext}=1/\ell$, and we must have $m\ge m_{\rm ext}$. We can easily see that
\be
M=\ft14 T + \fft{b}{\ell}\,,\qquad S-\ft18 \log(\ell^2\lambda)=\ft18
\log \Big(1 + \fft{\ell T}{2b}\Big)\,,\qquad
Q=32\pi^2b(2b + \ell T)\,.\label{MSQ}
\ee
These quantities are all nonnegative, indicating that they are all well defined in the allowed parameter region. (The thermodynamic constant contribution to the entropy above comes from the topological Einstein-Hilbert term.) However the entropy becomes divergent as $b\rightarrow 0$, which, as we shall see presently, implies that the free energy is unbounded below. It follows from \eqref{MSQ}, we have
\be
M^2 = \fft{Q}{64\pi^2\ell^2} + \fft{T^2}{16}\ge \fft{Q}{64\pi^2\ell^2}\,.
\ee
The extremal black hole saturates the inequality.  It is worth commenting here that if the parameter $m$ is less than $m_{\rm ext}$, then the function $f(r)$ in the metric has no real roots, and the metric describes a wormhole, connecting two AdS$_2$ boundaries at $r\rightarrow \pm\infty$. However, the full solution is actually singular since the scalar $\phi$ is singular at $r=a$. Further discussions on wormholes will be given in section \ref{sec:wh}.

According to the Euclidean action, the thermodynamic system is a canonical ensemble with Helmholtz free energy. The specific heat is given by
\be
c=T\fft{\partial S}{\partial T}\Big|_Q = \fft{1}{4(1 + \fft{4b}{\ell T})}>0\,.
\ee
This suggests that the black hole is locally stable, but the free energy tells a different story. There is a thermodynamic global instability associated with the free energy
\bea
F &=&\fft{b}{\ell} + \fft{T}{4} - \fft18 T \log\Big(1+ \fft{\ell T}{b}\Big)\nn\\
&=&\frac{\sqrt{Q+4 \pi ^2 \ell^2 T^2}}{8 \pi\ell } + \frac{T}{8}\log \left(\frac{Q}{\left(\sqrt{Q+4 \pi ^2\ell^2 T^2}+2 \pi\ell  T\right)^2}\right).
\eea
In this expression, we have chosen without loss of generality $\lambda \ell^2=1$, such that $\bar \phi=0$. For the given metric that is specified by the temperature $T$, the mass and free energy $(M,F)$ are both monotonously increasing functions of the scalar charge $Q$. For the constant-scalar black hole that sits at $b\rightarrow \infty$ ($Q\sim \infty$), both mass and free energy approach positively infinity. As $b\rightarrow 0$ ($Q\rightarrow 0$), we have $M\rightarrow T/4$, but the free energy is unbounded below with logarithmic divergence, unless the temperature vanishes. This suggests that the extremal black hole is stable, but non-extremal black holes are not. The time-evolution of the non-extremal black holes should lead to the vanishing of $e^{-2\phi}$ at the later time. We shall verify this in the next subsection.

\subsection{General time-dependent black holes}

Motivated by the thermodynamic global instability of the black holes, we construct the time-dependent solution here. By time dependence, we mean that the scalar field $\phi$ depends not only $r$, but also the time $t$. Define $\varphi = e^\phi$, we find that the Einstein equation in the $tr$ direction implies that
\be
\dot \varphi' - \fft{r^2-\ell^2 m}{\ell^2 f} \dot \varphi=0\,.
\ee
This equation can be completely solved, and substituting the result into the remainder of the scalar equations \eqref{geneom}, we find that the most general solution is given by
\be
\phi(r,t)=\log\Big(\alpha +\beta (r-m\ell^2) + \gamma\ell \sqrt{f(r)}\cosh\big(2\pi T (t-t_0)\big)\Big)\,.\label{phirt}
\ee
The scalar is well defined on and outside of the horizon where $f(r)\ge 0$. Inside the horizon where $f(r)<0$, we need to analytically continue to $t\rightarrow t + {\rm i}/(4T)$, and the scalar singularity can arise. The four parameters $(\alpha, \beta, \gamma, t_0)$ are not all independent. They satisfy the constraint
\be
\lambda\ell^2=\alpha^2 +4\pi^2T^2\,(\gamma^2 - \beta^2)\ell^4\,.
\ee
The static solution arises by setting $\gamma=0$ and redefining the parameters to be
$a=m\ell^2 - {\alpha}/{\beta}$. For the general solution with nonvanishing $\gamma$, the scalar field is unstable and becomes divergent at later time with $e^{-2\phi}\rightarrow 0$, confirming the global analysis in the previous subsection. The scalar has linear time dependence in later time, namely
\be
\phi \sim 2\pi T\, t\,,\qquad \hbox{as}\qquad t\rightarrow \infty\,.
\ee
There are two notable exceptions. One is when the black hole is extremal, in which case, the black hole is stable and time independent. This confirms the prediction in the thermodynamic analysis of the static black hole. The other is that for general black holes, the scalar on the horizon with $f(r_+)=0$ is also stable, independent of time. This is a characteristic feature of 2d gravities and is independent of coordinate transformation. We can thus describe a 2d black hole with the vacuum metric and the null line of the black hole is specified by the constant scalar field \cite{Almheiri:2014cka}.

\subsection{Including the Hamilton-Jacobi counterterm}

A Hamilton-Jacobi (HJ) counterterm was introduced in \cite{Grumiller:2007ju} for a general class of 2d gravities. For the theory \eqref{2dlag}, it takes the form \cite{Grumiller:2007wb} in the Euclidean signature signature
\be
S_{\rm HJ}=\fft{c}{8\pi} \int_{\rm \partial {\cal M}} dt \sqrt{|h|}\,\sqrt{\lambda} e^{-\phi}\,,\label{HJ}
\ee
where we add an additional dimensionless constant $c$, and it was determined in \cite{Grumiller:2007wb} as $c=-1$. In fact in the limit of $\phi=\bar \phi + \epsilon \varphi$ discussed in subsection 2.2, the bulk action of \eqref{totalaction0} becomes that of JT gravity, and the total boundary terms of \eqref{totalaction0} and \eqref{HJ} yields the one of JT gravity only when $c=1$. To be specific, we have
\be
S_{HJ}^{c=1} + \fft{1}{8\pi} \int_{\partial {\cal M}} dt \sqrt{|h|}(\phi K - \fft{1}{\ell})\sim \fft{\epsilon}{8\pi}\int_{\partial {\cal M}} dt\,\sqrt{|h|} \varphi (K-\fft{1}{\ell})\,.
\ee
The variation of this boundary action \eqref{HJ} simply vanishes under the Dirichlet boundary condition and therefore it will not affect the variation principle that yields the equation of motion. Furthermore, this term is convergent on the AdS$_2$ boundary even for a generic value of $c$, and its contribution to the free energy is $c \sqrt{f(a)}/(8\pi\ell)$. This implies that the HJ counterterm modifies the mass of the black hole to become
\be
M=\fft{\ell^2 m - a}{8\pi \ell^2} + \fft{c}{8\pi\ell}\sqrt{f(a)}\,.
\ee
It can be easily verified that the first law \eqref{firstlaw} continues to hold. Therefore we see that the HJ counterterm has no effect on the thermodynamic local properties. It can however affect the global properties. It can be established that the mass for the black holes are all nonnegative provided that $c\ge -1$ and the value determined by \cite{Grumiller:2007wb} saturates the bound. Since $f(a)$ vanishes as $a\rightarrow r_-$, it does not affect the conclusion that the free energy is unbounded below as the scalar charge $Q$ approaches zero.

\section{Wormholes}
\label{sec:wh}

\subsection{$\Lambda_0=-1/\ell^2>0$}

When the bare cosmological constant $\Lambda_0$ is positive, the 2d spacetime becomes locally dS$_2$, in which case, the two roots in the metric function $f$ describe the cosmic horizons rather than the black hole horizons. Therefore there can be no black hole in this case, but only the traversable wormhole that connects two cosmic horizons. It is more convenient to define $\tilde \ell^2 = -\ell^2>0 $ and $\tilde\lambda = -\lambda>0$. The metric and the scalar are now given by
\be
f=-\fft{r^2}{\tilde \ell^2} + 1 - 2m r\,,\qquad
\phi = \log\Big(\alpha + \beta (r + m \tilde \ell^2) + \gamma \tilde \ell \sqrt{f}\,
\cosh\Big(\fft{\sqrt{1-m^2\tilde \ell^2}}{\tilde\ell} (t-t_0)\Big)\Big)\,,
\ee
where the constants $(\alpha, \beta, \gamma, t_0)$ are subject to the constraint
\be
\tilde \lambda \tilde \ell^2 = \alpha^2 - (1 + m^2\tilde\ell^2) \tilde \ell^2 (\gamma^2 + \beta^2)\,.\label{whcond1}
\ee
The existence of dS$_2$ vacuum with costant $\phi$ implies that $\tilde \lambda>0$. The function $f$ is quadratic and always has two roots $r_\pm^c$, regardless the real value of $m$:
\be
r_\pm^c = -m\tilde \ell^2 \pm \tilde \ell \sqrt{1 + m^2\tilde \ell^2}\,.
\ee
The solution is static with $(r,t)$ being spacelike and timelike respectively in the region $r\in (r_-^c, r_+^c)$, confirming our earlier statement that the solution describes a traversable wormhole connecting the two cosmic horizons $r_\pm^c$, beyond which the spacetime becomes cosmological. The condition \eqref{whcond1} ensures that the scalar is free from singularity within the cosmic horizons; the singularity lies outside the cosmic horizons. The wormhole becomes symmetric if we set $\beta=0$, and can be made manifestly in the metric by the coordinate shift $r\rightarrow r - m\tilde\ell^2$.

\subsection{$\Lambda_0=-1/\ell^2<0$}

In this case, as we have discussed earlier, the general solution describes black holes for $m^2\ell^2\ge 1$. When $m^2\ell^2<1$, the metric becomes a traversable wormhole; however,
as we illustrated in subsection 3.1, there is no static scalar solution that is well defined in the whole wormhole space, except when the scalar is a constant. The conclusion remains true for time-dependent scalar solution; however, the singularity becomes rather intriguing when we allow the time dependence. For $1-m^2\ell^2>0$, the scalar solution, after taking $k=1$, is given by
\be
\phi= \log\Big(\alpha + \gamma \sqrt{f} \sin \big(\fft{\sqrt{1-m^2\ell^2}}{\ell}t\big)\Big)\,.
\ee
Note that we have set the corresponding $\beta$ in \eqref{phirt} to zero so that the ``wormhole'' solution is symmetric between $\tilde r$ and $-\tilde r$, where $\tilde r = r - m \ell^2$ and $f=\tilde r^2/\ell^2 + 1 - m^2/\ell^2$. The parameters $(\alpha,\gamma)$ are subject to the constraint
\be
\lambda \ell^2 = \alpha^2 - \gamma^2 (1-m^2\ell^2)\,.\label{whcond2}
\ee
The wormhole ``throat'' is located at $\tilde r=0$ for which the function $f$ is at its minimum $f_{\rm min}=1-m^2\ell^2$. The condition \eqref{whcond2} ensures that $\alpha^2 - \gamma^2 f_{\rm min} >0$.

The scalar function is periodic with period $\Delta t=\tau_0 \equiv 2\pi\ell/\sqrt{1-m^2\ell^2}$. Intriguingly, during half of the period, namely $t\in [0, \fft12\tau_0]$, the wormhole opens up, and the solution is free from spatial singularity and it connects two asymptotic AdS$_2$ boundaries. The situation changes in the other half, $t\in (\fft12\tau_0, \tau_0)$, and the two scalar singularities enclose the spacetime from reaching the boundaries. The space during this half period is confined in
\be
|\tilde r| \le \ell \sqrt{ \fft{\alpha^2}{\gamma^2 \sin^2\big(\fft{2\pi}{\tau_0} t\big)} - f_{\rm min}}\,.
\ee
As time $t$ increases from $\fft12\tau_0$, the world size decreases monotonously from infinity to the minimum $|\tilde r|<\tilde r_{\rm min}=\ell \sqrt{\alpha^2/\gamma^2 - f_{\rm min}}$ at $3\tau_0/4$. The condition \eqref{whcond2} ensures that a real value of $\tilde r_{\rm min}$ always exists. The world increases its size as $t$ runs from $3\tau_0/4$ to the whole wormhole space at $t=\tau_0$, after which the wormhole fully opens up for the next half period, and then repeats the whole time cycle.

\section{Euclidean AdS$_2$ and the action}
\label{sec:h2act}

We would now like to study the effective theory of the AdS$_2$ boundary. We consider Euclidean signature and the AdS$_2$ vacuum is hyperbolic 2-space
\be
ds_2^2 = \fft{dz^2 + dt^2}{z^2}\,.\label{ads2}
\ee
Without loss of generality, we have set $\ell=1$. The most general scalar solution in this background is
\be
\phi=\log\Big(\fft{\alpha + \beta (z\cos\delta + t\sin\delta) + \gamma (t^2 + z^2)}{z}\Big)\,,
\ee
together with the constraint on the constant parameters $(\alpha, \beta,\gamma, \delta)$:
\be
\lambda=\beta^2 - 4\alpha \gamma\,.
\ee
The vacuum solution where $\phi$ is a constant corresponds to taking all $(\alpha, \gamma, \delta)$ to be zero, in which case $\lambda=\beta^2$. Therefore we must have $\lambda>0$. When the scalar is constant, the AdS$_2$ vacuum is dual to CFT$_1$, which is vacuous and trivial. The general scalar solution breaks the AdS$_2$ isometry to become nearly AdS$_2$ and the dual theory is the NCFT$_1$, as in the case of the JT gravity/SYK correspondence.

In this section we shall study the total action
\be
S=\int_{\cal M} d^2x\,\sqrt{g} \Big(\phi R +\fft{1}{\ell^2} -\lambda\, e^{-2\phi} - (\nabla\phi)^2\Big) + 2\int_{\partial {\cal M}} dy \sqrt{h}(\phi K - \fft{1}{\ell})\,,\label{totalaction}
\ee
where the $h$ is the induced metric on the relevant boundary. The first term is the bulk action $S_{\rm bulk}$ and the second term is the surface action $S_{\rm surf}$ involving both Gibbons-Hawking-York term and the holographic counterterm.

\subsection{Boundary slices}

The boundary of the Euclidean AdS$_2$ space \eqref{ads2} is at $z=0$, and the effective action from $z=\infty$ to $z=0$ is unremarkable. Nontrivial results can be obtained by considering new slices of the original boundary, parameterized by a new time coordinate. We make a coordinate transformation
\be
t=t(u)\,,\qquad z=z(u,\epsilon)\,,\label{gencoord}
\ee
where $u$ is understood to be the new time and $\epsilon$ is the new radial variable. Thus we have
\be
ds^2 = g_{uu} du^2 + g_{\epsilon\epsilon} d\epsilon^2 + 2g_{u\epsilon} du d\epsilon\,,
\ee
where
\be
g_{uu}=\fft{\dot t^2 + \dot z^2}{z^2}\,,\qquad
g_{\epsilon\epsilon}=\fft{z'^2}{z^2},,\qquad g_{u\epsilon}=\fft{\dot z z'}{z^2}\,.
\ee
Here a dot denotes a derivative with respect to $u$ and a prime a derivative with respect to $\epsilon$. The extrinsic curvature on the constant $\epsilon$ surface is
\be
K=\fft{\dot t^3 - z \dot z\,\ddot t + \dot z^2 \dot t
-z\ddot z \dot t}{(\dot t^2 + \dot z^2)^{3/2}}\,.
\ee

Following \cite{Maldacena:2016upp}, we would like to require that $g_{uu}=1/\epsilon^2$, at the boundary $\epsilon\rightarrow 0$. This can be achieved by making an ansatz
\be
z=\epsilon z_1(u) + \epsilon^3 z_3(u) + \epsilon^5 z_5(u) + \cdots\,,\label{ansatz}
\ee
which implies
\be
g_{uu}=\frac{\dot t^2}{z_1^2 \epsilon^2}+\frac{z_1 \dot z_1^2-2 z_3 \dot t^2}{z_1^3}+\frac{\epsilon ^2 \left(-2 z_5 z_1\dot t^2+3 z_3^2 \dot t^2+2 z_1^2 \dot z_1 \dot z_3-2 z_3 z_1\dot z_1^2\right)}{z_1^4}+O\left(\epsilon ^3\right)\,.
\ee
We can solve $z_1,z_3,z_5$ etc., order by order, and we have
\be
z_1=\dot t\,,\qquad z_3=\fft{\ddot t\,^2}{2\dot t}\,,\qquad
z_5=\fft{8 \dot t\, \ddot t\,^2 \dddot t - 5 \ddot t^4}{8\dot t^3}\,.\label{zis}
\ee
Thus, for small $\epsilon$ expansion, we have
\bea
K &=& 1 + \epsilon^2 K_2 + \epsilon^4 K_4 + \cdots\,,\qquad K_2 = -\frac{2 z_1 \dot z_1 \ddot t+\dot t \left(\dot z_1^2-2 z_1 \ddot z_1\right)}{2 \dot t^3}\,,\nn\\
K_4 &=& \fft{1}{8\dot t^5}\Big(12 z_1 \dot z_1^3 \ddot t -
8(z_3 \dot z_1 + z_1 \dot z_3)\dot t^2 \ddot t +
3\dot t (\dot z_1^4 - 4 z_1 \dot z_1^2 \ddot z_1) +
8\dot t^3 (z_1\ddot z_3 + z_3 \ddot z_1 - \dot z_1 \dot z_3)\Big)\,.
\eea
Thus we see that $K_2$ is determined solely by $z_1$ and $K_4$ is determined by $(z_1, z_3)$. For $z_i$'s given by \eqref{zis}, we have
\bea
K_2 &=& \fft{2\dot t\, \dddot t - 3 \ddot t^2}{2\dot t^2}\equiv {\rm Sch}(t(u),u)\,,\nn\\
K_4 &=& \fft{8\dot t^2\, \ddot t\,\, \ddddot t + 8 \dot t^2\, \dddot t\,^2 - 44 \dot t\,\ddot t\,^2\,\dddot t + 27 \ddot t^4}{8\dot t^4}\,.
\eea
Note that the $K_2$ term is precisely the Schwarzian derivative, which is invariant under the general $GL(2,\mathbb R)$ transformation
\be
t\rightarrow \fft{a t + b}{c t +d}\,.
\ee
In this new boundary slices, the scalar field is given by
\bea
\phi &=& \log\Big(\fft{1}{\epsilon} (\psi_0 + \epsilon \psi_1 + \psi_2 + \cdots)\Big)\,,\nn\\
\psi_0 &=& \fft{\alpha + \beta\sin\delta\, t + \gamma\, t^2}{\dot t}\,,\qquad \psi_1= \beta \cos\delta\,,\qquad \cdots\,,\label{psi0psi1}
\eea
where the ellipsis denotes the higher-order terms in the $\epsilon$ expansion.

\subsection{The on-shell action}

For JT gravity, the bulk term in the action vanishes identically, thus only the surface term contributes to the total action. This leads immediately to the Schwarzian action \cite{Maldacena:2016upp}
\be
S_{\rm sch} = 2 \int du\, \Phi_{\rm inf}\, {\rm Sch}(t(u),u)\,.\label{schact}
\ee
In our case, the bulk contribution does not vanish, given by
\bea
S_{\rm bulk}(\epsilon) &=& -\int du\Bigg(\fft{2\dot t}{z} (\phi-1 - \fft{\lambda(\beta\cos\theta + 2 \gamma z)}{\chi^2-\lambda} e^{-\phi})\nn\\
 &&\qquad- \fft{8 \dot t \gamma\chi^2}{(\chi^2-\lambda)^{3/2}} \Big(\arctan
\Big(\fft{\beta\cos\theta + 2\gamma z}{\sqrt{\chi^2-\lambda}}\Big)-\ft12\pi\Big)\Bigg)\Big|_\epsilon\,,
\eea
where $\chi=\beta \sin\theta + 2 \gamma t$. Note that we have integrated the radial variable $\epsilon$ from $\infty$ to $\epsilon\rightarrow 0$. The asymptotic boundary of the AdS$_2$ is located at $\epsilon=0$, but we would like to consider the finite cutoff away from the boundary with $\epsilon$ as an expansion parameter.  Together with the surface contribution, the total on-shell action is given by
\bea
S_{\rm tot} = \int du (L_{\epsilon} + \dot Y_\epsilon)\,,
\eea
with
\be
L_\epsilon = L_0 + \epsilon L_1 + \epsilon^2 L_2 + \epsilon^3 L_3 + \cdots\,,\qquad
Y_\epsilon=\log\epsilon\Big(\epsilon \widetilde Y_1 + \epsilon^3 \widetilde Y_3 + \cdots\Big) + \epsilon Y_1 + \epsilon^3 Y_3 + \cdots\,.
\ee
Here the $Y_\epsilon$ terms are total derivative and do not contribute the dynamics. Up to and including the $\epsilon^3$ order, we find
\bea
L_0 &=& \fft{2\lambda\psi_1}{\psi_0(\chi^2-\lambda)} -
\fft{4\gamma \chi^2(\pi -\arctan(\fft{\psi_1}{\sqrt{\chi^2-\lambda}}))}{
(\chi^2-\lambda)^{3/2}}\dot t\,,\nn\\
L_1 &=& -2 {\rm Sch} + \fft{2\psi_1^2}{\psi_0^2}\,,\qquad
L_2=\fft{2\psi_1}{\psi_0} \Big(\fft{d}{du}\big(\fft{\ddot t}{\dot t}\big)-\fft{\gamma \dot t}{\psi_0} - \fft{\lambda}{\psi_0^2}\Big),\nn\\
L_3 &=& - \fft23 K_4 +\frac{2 \chi  \ddot t\,^3}{3 \psi _0 \dot t^3}
-\frac{\left(\chi ^2-2 \psi _1^2-\lambda\right)\ddot t\,^2 }{\psi _0^2\, \dot t^2}
+\frac{\chi ^4-8 \chi ^2 \psi _1^2+3 \psi _1^4 - 2\lambda(\chi^2-4\psi_1^2) +\lambda^2 }{6 \psi _0^4}\,,\nn\\
\widetilde Y_1 &=& -\fft{2\ddot t}{\dot t}\,,\qquad
\widetilde Y_3 = \fft{5\ddot t\,^3}{3\dot t^3} - \fft{2 \ddot t\,\, \dddot t}{\dot t^2}\,,\qquad
Y_1 = -\fft{2\chi}{\psi_0} + 2(1 + \log\psi_0)\fft{\ddot t}{\dot t}\,,\nn\\
Y_3 &=& \frac{\chi  \left(\chi ^2-3 \psi_1^2-\lambda\right)}{6 \psi_0^3}
+\frac{\left(\chi^2 -\psi _1^2-\lambda\right)\ddot t}{2 \psi _0^2 \dot t}-
\frac{\chi  \ddot t\,^2}{\psi _0 \dot t^2} -
\fft{5 (2+3\log\psi_0) \ddot t\,^3}{9 \dot t^3}\nn\\
&& +
\fft{2 (1 + 3\log \psi_0)\ddot t\,\,\dddot t}{3\dot t^2}\,.
\eea
The on-shell action in general is rather complicated. However, it can be easily established that $L_0$ is a total derivative, since it takes the form of $L_0=U(t(u)) \dot t$, for some complicated function of $U$. That $L_0$ is trivial may be related to the fact that the theory admits full AdS$_2$ vacuum whose CFT dual is trivial. Nontrivial dynamics emerges only at the finite cutoff away from the boundary, associated with the higher-order of $\epsilon$ terms, where the full conformal symmetry is broken.

The action simplifies dramatically in later time $t\rightarrow \infty$, for which, $\psi_0\sim t^2$ and $\chi\sim t$. We therefore have
\be
L_{\epsilon}\Big|_{t\rightarrow \infty}= -2 {\rm Sch}\,\epsilon - \fft23 K_4\,\epsilon^3 + \cdots\,.\label{latertime}
\ee
Thus we see that at the leading $\epsilon$ order, the later time dynamics is governed by the Schwarzian action, at the finite cutoff of the AdS$_2$ boundary.

We now consider the effect of the Hamilton-Jacobi counterterm \eqref{HJ} to the on-shell action. It is convergent and its contribution to the total action is
\bea
\fft{8\pi}{c\sqrt{\lambda}} L_\epsilon^{\lambda_{\rm B}} &=& \fft{1}{\psi_0} -
\fft{\psi_1}{\psi_0^2} \epsilon +
\Big(\frac{\psi _1^2}{\psi _0^3}-\frac{\gamma  t'}{\psi _0^2}+\frac{t''\,^2}{2 \psi _0 t'^2}\Big)\epsilon^2 - \fft{\psi_1}{\psi_0^2} \Big(\frac{\psi _1^2}{\psi _0^2}-\frac{2 \gamma  t'}{\psi _0}+\frac{t''\,^2}{t'^2}\Big) \epsilon^3 + \cdots
\eea
The leading term is again a total derivative, and furthermore these terms all vanish at later time and hence do not affect the conclusion \eqref{latertime}.

\subsection{The semi-on-shell action}

In the case of JT gravity, the $\Phi_{\rm inf}(u)$ in \eqref{schact} is the background field and the equation motion associated with the variation of $t(u)$ can be solved precisely by the on-shell $\Phi_{\rm inf}$ that can be also obtained from the general bulk solution. A coordinate transformation $\Phi_{\rm inf} du=d\tilde u$ is necessarily so that the action in \eqref{schact} become Schwarzian, where $\Phi_{\rm inf}$ can be treated as a constant \cite{Maldacena:2016upp}.

Here we would like to repeat the same analysis and treat the $(\psi_0,\psi_1,\psi_2)$ {\it etc.}~as abstract background field and rederive the action.  We shall obtain the action only up to and including the $\epsilon$ order, therefore, we only need to consider
\be
z=\epsilon\, \dot t\,,\qquad \log\Big(\fft{1}{\epsilon} \psi_0(u) + \psi_1(u) + \epsilon\psi_2(u)\Big)\,.
\ee
The convergence of the bulk action at $\epsilon\rightarrow \infty$ requires that $\psi_2=\gamma \dot t$, where for convenience we have chosen the same $\gamma$ parameter as in the on-shell solution.  We find that total action is given by
\bea
L_0 &=& \fft{1}{\gamma\psi_0 \dot t^2\, (4\gamma \psi_0 \dot t- \psi_1^2)}
\Big(\gamma  \psi _0^2 \psi _1 \ddot t\, ^2 + 2 \gamma\psi _0\,\dot t\,\ddot t \left(\psi _1 \dot \psi _0-2 \psi _0 \dot \psi _1\right) +\psi _0 \psi _1 \dot t \dot \psi _1^2\nn\\
&& +\gamma  \dot t^2 \left(\psi _1^3 +\psi _1 \left(\lambda +\dot \psi _0^2\right)-4 \psi _0 \dot \psi _0\dot \psi _1\right)  -4 \gamma ^2 \psi _0 \psi _1 \dot t^3
\Big)\nn\\
&&+\fft{2}{\dot t (4\gamma \psi_0 \dot t- \psi_1^2)^{3/2}}\Big(\pi - 2 \arctan\big(\fft{\psi_1}{\sqrt{4\gamma \psi_0 \dot t- \psi_1^2}}\big)\Big)\Big(\gamma \psi_0^2
\ddot t\,^2 + 4 \gamma ^2 \psi _0 \dot t^3\nn\\
&&+\psi _0 \ddot t \left(2 \gamma  \dot t \dot \psi _0-\psi _1 \dot \psi _1\right) +
\dot t \dot \psi _1 \left(\psi _0 \dot \psi _1-\psi _1 \dot \psi _0\right) +
\gamma  \dot t ^2 \left(\lambda +\dot \psi _0^2-\psi _1^2\right)
\Big)\,.
\eea
Note that now the fields $\psi_0$ and $\psi_1$ are the abstract background fields. Although this is a rather complicated Lagrangian; nevertheless, we find that its equation of motion associated with the variation of the dynamic variable $t$, namely
\be
\fft{\partial L_0}{\partial t} - \fft{d}{du}\left(\fft{\partial L_0}{\partial \dot t}\right) +
\fft{d^2}{du^2}\left(\fft{\partial L_0}{\partial \ddot t}\right)=0\,,\label{lagrange}
\ee
can be solved precisely by the on-shell solutions of $(\psi_0,\psi_1)$, given in \eqref{psi0psi1}. Note that the HJ counterterm contributes $1/\psi_0$ in the leading term, and it does not affect on the Lagrange equation \eqref{lagrange}. Thus as in the case of the black hole thermodynamics, the HJ term \eqref{HJ} does not affect our physical conclusion when the metric is AdS$_2$ or asymptotic to AdS$_2$.

At the $\epsilon\log\epsilon$ order, the action is a total derivative $\dot {\widetilde Y_1}$, independent of $(\psi_0,\psi_1)$, as in the case of the on-shell action. At the $\epsilon$ order we have
\be
L_1 = \fft{\lambda}{\psi_0^2} + \fft{\psi_1^2+\dot \psi_0^2}{\psi_0^2} + \fft{d}{du} \Big(2\log(\psi_0)\,\fft{\ddot t}{\dot t}\Big)\,.
\ee
Substituting the on-shell results obtained from $L_0$ yields precisely the on-shell action obtained earlier at this order. We therefore achieved a consistent picture of the boundary dynamics of the nearly AdS$_2$ at the finite cutoff.

\section{Euclidean dS$_2$ and its action}
\label{sec:s2act}

When the bare cosmological constant is positive, corresponding to, without loss of generality, $\ell^2=-1$, the metric is dS$_2$. In Euclidean signature, it is a unit $S^2$
\be
ds_2^2 = d\theta^2 + \sin^2\theta\, d\tau^2\,,
\ee
with $\theta \in [0,\pi]$ and $\tau \in [0,2\pi)$. The most general solution for the scalar field is
\be
\phi = \log\Big(\alpha + \beta \cos\theta + \gamma \sin\theta\,\cos(\tau-\tau_0)\Big)\,,
\ee
with the parameter $\lambda$ given by
\be
\lambda =  \beta^2 +\gamma^2-\alpha^2\,.
\ee
The vacuum, where $\phi$ is a constant, emerges when $\beta=0=\gamma$, and hence $\lambda=-\alpha^2$. Thus we must have $\lambda <0$ for the theory to admit such a vacuum.
For the scalar to be well defined on the 2-sphere, we must have
\be
\alpha>\sqrt{\beta^2 + \gamma^2}\,,
\ee
which is precisely satisfied by $\lambda<0$. It is worth comparing to JT gravity with a positive cosmological constant, in which case, the scalar is given by
\be
\varphi =\alpha \cos\theta + \beta \sin\theta\,\cos(\tau-\tau_0)\,,
\ee
where $(\alpha, \beta, \tau_0)$ are free parameters. Thus we see that the $\varphi=0$ singularity cannot be avoided on the manifold. We find that the on-shell action for the $S^2$ solution is
\bea
S &=& 4\pi \Big(2\sqrt{\fft{\alpha^2-\beta^2-\gamma^2}{\alpha^2-\beta^2}}-2+\log(\alpha^2-\beta^2)\Big)\nn\\
&& + 4\gamma^3\int_{0}^{2\pi} d\tau \fft{\cos\tau \sin^2\tau}{(\alpha^2-
\beta^2 -\gamma^2 \cos^2\tau)^{3/2}} \arctan
\Big(\fft{\gamma \cos\tau}{\sqrt{\alpha^2-\beta^2-\gamma^2\cos^2\tau}}\Big)\nn\\
&=& 4\pi \log(-\lambda).
\eea
(The above provides a nice formula for the definite integration in the second line.) In other words, the parameters in the scalar field does not affect the action. Note that since $S^2$ has no boundary and the scalar is well defined, there is no boundary contribution in the evaluation of the action.

\section{Conclusion}

In this paper, we studied a candidate theory for Einstein gravity at the $D\rightarrow 2$ limit. The theory was obtained from Kaluza-Klein reduction of Einstein gravity in general $D=2+n$ dimensions, reduced on $n$-dimensional internal Einstein space of constant curvature $\lambda$, keeping only the breathing scalar mode. Under some suitable $n\rightarrow 0$ limit, a scalar-tensor theory involving Ricci scalar emerges. The theory was known to be related to Liouville CFT of a large central charge. We found that when the theory is minimally coupled to a bare cosmological constant, the theory can also reduce to JT gravity that describes the perturbative dynamics around the scalar fixed point.

We studied the properties of the theory by constructing the full solution space. We first construct the static black holes, which is governed by two parameters associated with the mass $M$ and the charge $Q$ of scalar hair. We obtained the first law of black hole thermodynamics and derived from the Euclidean action that thermodynamic potential is the Helmholtz free energy.
Our black hole mass formula indicate that the AdS$_2$ vacuum at the scalar fixed point has infinity mass and free energy, indicating that it is not a stable vacuum.

We found that the black holes were all locally stable since the specific heat is nonnegative. However, for non-extremal black holes, there is a thermodynamic global instability associated with the scalar charge.  Specifically, the black hole metric is specified by temperature $T$ only. For the scalar field to be absent from a singularity, the scalar charge $Q$ must be non-vanishing. As $Q\rightarrow \infty$, the solution becomes the black hole with the constant scalar, and the mass and free energy are both positively infinite. As $Q\rightarrow 0$, the mass is finite $M\rightarrow T/4$ and the free energy becomes unbounded below, except at zero temperature.  This motivated us to construct the most general time-dependent black hole and we saw that the time-dependence dropped out at zero temperature, indicating that extremal black holes are stable.  When $T>0$, the time evolution drives the black hole to the singular $Q\rightarrow 0$ limit. The 2d gravity model provides us a simple analytical example of illustrating the connection between the black hole thermodynamic instability and its time evolution.

It is straightforward to construct traversable wormhole metrics in two dimensions that connect two asymptotic AdS boundaries. However, we found that scalar field was necessary singulary. However, we showed that this singularity has intriguing properties when the scalar was allowed to be time dependent and it was periodic. The wormhole opens up and connects the two AdS$_2$ boundaries during half of the period, and singularities develop and close in during the other half. When the cosmological constant is positive and the metric is asymptotic to dS$_2$, the solution is necessarily a traversable wormhole connecting two cosmic horizons. In this case, the scalar singularities lie beyond the cosmic horizons.

We further studied the solution space in the Euclidean signature. For the negative bare cosmological constant, the 2-space is hyperbolic. We followed the work of \cite{Maldacena:2016upp} and studied the boundary dynamics in some nontrivial boundary slices. We found that the total action was trivial in that it is a total derivative, perhaps consistent with the fact that the theory has the full AdS$_2$ fixed point. Nontrivial dynamics could arise if we moved away from the asymptotic boundary to some finite boundary slices. The Schwarzian action with $SL(2,{\mathbb R})$ symmetry would arise at the leading order of the finite boundary cutoff, but only at later time with $t\rightarrow \infty$. For a positive bare cosmological constant, the space is a 2-sphere, and we found that the scalar field was well defined everywhere on the 2-manifold. This should be contrasted to JT gravity where the scalar would suffer some inevitable singularities. To conclude, the properties we uncovered in this paper make the theory a particularly interesting model to investigate further.

\section*{Acknowledgement}

We are grateful to Yue-Zhou Li for many useful discussions. This work was supported in part by NSFC (National Natural Science Foundation of China) Grants No.~11875200 and No.~11935009.

\end{document}